\documentclass[sigconf,natbib=true,table]{acmart}
\pdfoutput=1
\AtBeginDocument{%
  \providecommand\BibTeX{{%
    \normalfont B\kern-0.5em{\scshape i\kern-0.25em b}\kern-0.8em\TeX}}}

\usepackage{dirtytalk}
\usepackage{subcaption}
\usepackage{bbm}
\usepackage[export]{adjustbox}
\usepackage{multirow}
\usepackage{makecell}

\setcopyright{none}
\renewcommand\footnotetextcopyrightpermission[1]{} 
\begin{document}

\title{Not Just Skipping: Understanding the Effect of Sponsored Content on Users' Decision-Making in Online Health Search}


\author{Anat Hashavit$^1$, \ Hongning Wang$^2$,  \  Tamar Stern$^1$, \  Sarit Kraus$^1$}
\affiliation{%
  \institution{$^1$Bar Ilan University,  $^2$University of Virginia}
  \country{$^1$Ramat~Gan, Israel, $^2$Charlottesville, VA, USA}}
\email{{anat.hashavit,stern.tamar96}@gmail.com, sarit@cs.biu.ac.il, hw5x@virginia.edu}

\renewcommand{\shortauthors}{Hashavit et al.}

\newcommand\hnote[1]{\textcolor{red}{[Wang]: #1}}
\begin{abstract}
  Advertisements (ads) are an innate part of search engine business models.
  Marketing research shows sponsored search advertisement is an effective tool to promote sales:
  advertisers are willing to pay search engines to promote their content to a prominent position in the search result page (SERP). 
  This raises concerns about the search engine manipulation effect (SEME): the opinions of users 
  can be influenced by the way search results are presented.
  
  In this work, we investigate the connection between SEME and sponsored content in the health domain. 
   We conduct a series of user studies in which participants need to evaluate the effectiveness of different non-prescription natural remedies for various medical conditions. 
   We present participants SERPs with different intentionally created biases towards certain viewpoints, with or without sponsored content, and ask them to evaluate the effectiveness of the treatment only based on the information presented to them.
   We investigate two types of sponsored content:
   1). Direct marketing ads that directly market the product without expressing an opinion about its effectiveness; and 2). Indirect marketing ads that explicitly advocate the product's effectiveness on the condition in the query. 
   Our results reveal a significant difference between the influence on users from these two types of ads. 
   Though direct marketing ads are mostly skipped by users, they do sometimes tilt  users decision making towards more positive viewpoints.  Indirect marketing ads affect both the users' examination behaviour and their perception of the treatment's effectiveness. We further discover that the contrast between the indirect marketing ads and the viewpoint presented in the organic search results plays an important role in users' decision-making. When the contrast is high, users exhibit a strong preference towards a negative viewpoint, and when the contrast is low or none, users exhibit preference towards a more positive viewpoint. 
   
   
   
  
 

\end{abstract}



\maketitle
\pagestyle{empty}

\begin{figure}[t]
\centering
\includegraphics[scale=0.35,keepaspectratio]{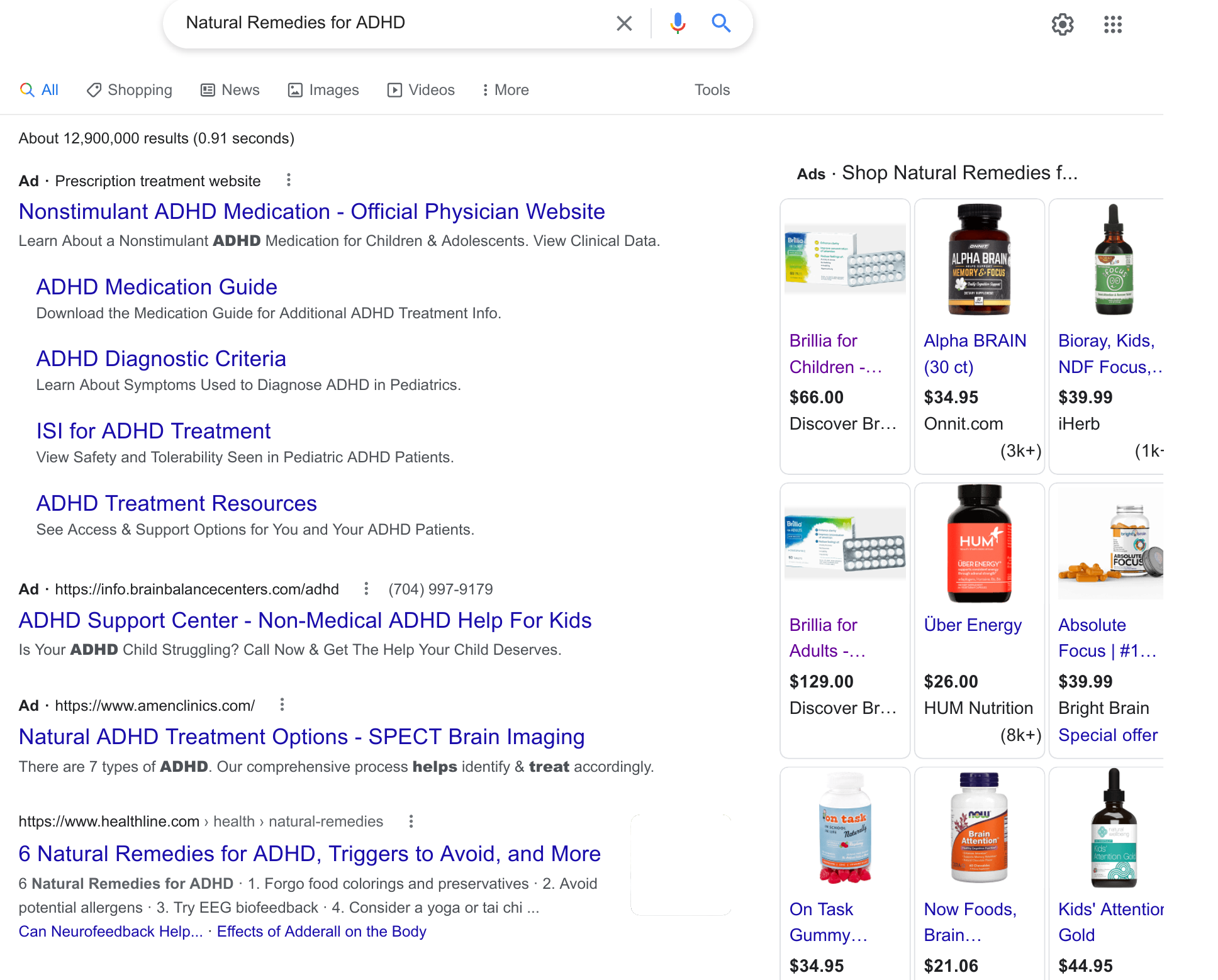}
\caption{Results for the query: `Natural remedies for adhd'. }
\label{fig:example}
\vspace{-3mm}
\end{figure}

\section{Introduction}


Search engines are a gateway to information, which help users make informed decisions in various aspects of their lives. This includes decisions about their health ~\cite{PEW}.
Users trust that search engines will present them with accurate and reliable results~\cite{PEW_trust}. However, this trust can make them susceptible to biases in the search results: 
whether it is the general and often positive bias of the presented results
~\cite{white2014content, white2013beliefs}; or the users' own tendency to inspect mostly the top results on the page 
~\cite{joachims2017accurately,keane2008people,o2006modeling,pan2007google}. In the health domain in particular, it has been shown that users are mainly susceptible to positive bias when researching medically related topics  ~\cite{allam2014impact,pogacar2017positive,white2013beliefs,hashavit2021understanding}. This problem can be exacerbated when financial and marketing influences come into play, if, for example, people count on natural remedies or vitamins as an alternative to professional medical treatment or vaccinations. This is a concerning phenomenon especially now during the COVID pandemic, when consuming inaccurate medical information can be life threatening.


On the other hand, search engines are also a powerful advertising platform since the advertisers can connect their content to relevant queries, thus specifically targeting relevant audiences~\cite{varian2006economics}. Ads are usually presented at the top of the search result page (SERP), ensuring that they are visible to users.
Search engine advertising has become very common and users are accustomed to encounter ads in their SERP. As a result, they sometimes develop a tendency to disregard the ads; skipping over them as if they do not exist, a phenomenon often referred to as banner blindness~\cite{li2019user,foulds2021investigating,burke2005high}.
Nevertheless, marketing research shows search engine advertising is an effective tool advancing various consumer metrics, such as brand awareness and brand consumption, even among users who do not click on the ads \cite{searchAdsEffect}.

In this work we focus on healthcare related search engine advertising. 
As sponsored content can introduce additional bias to the presented results, it can present a potential issue when the search process is part of a user's decision-making process in health-related problems. 
Take for example a parent who has a child with attention deficit hyperactive disorder (ADHD)
and wishes to explore options of natural-based remedies for treatment of ADHD.
The parent starts their research with their trusted search engine. Figure  \ref{fig:example} shows the top of the result page for the query ``natural remedies for ADHD''.
As can be seen, the first viewport of the SERP is almost completely dominated by sponsored content. 
The vast majority of the sponsored content is highly relevant to ADHD and 
promotes natural non-prescription products, whose names strongly suggest benefits to attention and focus.
Unfortunately, the fact is that most of these products have not been scientifically proven to be effective \cite{gillies2012polyunsaturated,heirs2007homeopathy}, but this information is not presented next to the ad. Although this information might be included in one of the returned results lower on the page, the parent might not notice them as extra effort is needed to examine those positions \cite{yue2010beyond,joachims2017accurately}. This bias can prevent the parent from making an informed decision that considers all relevant information, which in turn can harm their child ~\cite{harpin2016long}.\footnote{The effective course of treatment for ADHD is out of the scope of this paper.}

In this work, we explore the effect of sponsored content in search engines on users' decision-making in health related topics, in particular, under the task setting of determining the effectiveness of online available natural remedies to related medical conditions. 
We focus our study on the following research questions: 
\begin{itemize}
    \item \textbf{RQ1:}  How is users’ result examination behaviour affected by the sponsored content on a SERP?
    \item \textbf{RQ2:} When users interact with a search engine to decide about the effectiveness of a medical treatment, does the existence of sponsored content affect their decisions?
   
\end{itemize}

In order to answer these research questions, we conducted a series of user studies in which participants were presented with SERPs discussing the effectiveness of natural remedies on corresponding medical conditions. Participants were presented with a SERP that only included results that specifically discuss the related query but with different viewpoints (e.g., for vs. against).
The participants were then requested to only use the results presented to them to decide the effectiveness of the remedy mentioned in the query. Each SERP contained an equal number of retrieved results, but the order in which the results with different viewpoints were presented varied so that participants were exposed to different levels of rank bias. 
In the meanwhile, the participants could be presented with a SERP that did not contain any ads, or with an ad that directly markets the remedy discussed in the query, or with an indirect marketing  ad. In particular, indirect marketing is a marketing method in which instead of directly promoting a product or offering it for purchase, the advertisers create promotional content discussing the benefit of the product for the purpose of building the brand. 

We tracked the participants' click behaviour and their responses to our post-study survey questions and compared the results under various settings. 
We found that participants tended to skip direct marketing ads and proceed to organic content. However, the existence of the ads could still affect users' decision making and introduce positive bias. 
For indirect marketing ads, we found that the effect of ads on both participants' examination behaviour and their responses depended on the ranking bias of the organic results. When the organic results were biased towards a negative viewpoint, participants not only skipped the indirect marketing ads during their search but also enforced their negative bias towards the treatment's effectiveness. 
When the organic results were biased towards an inconclusive or positive viewpoint, however, participants were more inclined to inspect the content of the ads which introduced positive bias to their decision making. 



\section{Background and Hypothesis}
We explore the connection between search engine manipulation effect and sponsored content in the health search domain. In this section, we first discuss existing work related to all these aspects
and then present the null hypotheses related to our research questions.

\noindent\textbf{Impact on Users' Beliefs.} Most studies that explored search engines' effect on users' beliefs and decision-making show that higher ranked results are more likely to affect the users, especially  when the results are biased towards a positive viewpoint~\cite{allam2014impact,pogacar2017positive,epstein2015search,white2014belief}.
One exception to this is ~\cite{draws2021not} who examined the effect of position bias on debatable topics. This study examined the attitude change exhibited in users who held a neutral viewpoint before the study. Although users were affected by the content they read, the data collected by the study showed that despite \textit{position bias}, the order in which results were ranked did not have a significant effect on users' attitude change.

White and Hassan \cite{white2014content} showed that the search results in medical  related queries are often biased toward positive outcomes. White \cite{white2013beliefs} investigated the issue of bias in yes-no question based on online search results about medical conditions. Similar results are observed, showing that users favored more  positive results (irrespective of the truth). More so, White showed that almost half of the times the viewpoints presented in the search results were scientifically incorrect.
Pogacar et al. \cite{pogacar2017positive} presented users with biased SERP and showed that this bias affected users' decision-making in health related questions, and Hashavit et al. ~\cite{hashavit2021understanding} showed that this phenomena repeats even when the users are free to conduct the search on their own.

\noindent\textbf{Impact on Users' Search Experience.} Advertisements' effect on users' search experience has been researched in several studies. Lewandowski at el.
~\cite{lewandowski2017users, lewandowski2018empirical} conducted two large-scale studies on  German internet users that focused on {Google}. Their studies found that users had limited knowledge regarding a search engine's business model. Many of them reported that they were either unaware that it was possible to pay {Google} to present content on the SERP or they did not know how to distinguish between ads and organic results. 
Schulthei{\ss} and Lewandowski~\cite{schultheiss2021users} showed that users with less understanding of search engine business models were more likely to click on ads.
Several studies found that the quality of ads and their rank affected the attention that ads received as well as the probability the user would click on them~\cite{giraldo2021influence,alanazi2020impact,buscher2010good,phillips2013ads, foulds2021investigating,danescu2010competing,lagun2016understanding}. This correlates with many studies showing users exhibit strong \textit{position bias} towards higher ranked results ~\cite{joachims2017accurately,keane2008people,o2006modeling,pan2007google}. 
However, when examining attention to ads compared to organic results, studies found negative bias toward the sponsored content. Sponsored content received significantly less attention from users in comparison to organic results, and on several occasions they distracted and frustrated the users ~\cite{giraldo2021influence,buscher2010good,foulds2021investigating, danescu2010competing}.
A recent study by Foulds et. al ~\cite{foulds2021investigating} 
found that overall ads had a negative impact on users' search experience, their performance in the search task, and their search experience. 

Much research attention has be paid on sponsored content's effectiveness under various conditions from a marketing perspective. The goal of these studies is mainly to examine factors that make advertisement more effective. The main metric inspected is click through rate \cite{jansen2006examination, gauzente2010intention}, some inspect users' attitude towards online advertising \cite{lu2017sponsored} and percentage of successful purchases originated from ads \cite{agarwal2015organic}. These studies focus mainly on e-commerce  and do not explore the effect of sponsored content on users' beliefs.  

To briefly summarize, existing studies examined the effect of ads on users' behaviour and search experience from multiple angles but they did not touch the effect of sponsored content on users' perception or their beliefs. 
Our research focuses on not only the effect that ads have on users' search behavior, but also the effect that ads might have on users' decision-making, their beliefs and biases, and their perception of truth in relation to their information need. Previous work in information retrieval suggests users are prone to ignore the ads; but if the users did not pay attention to the search ads, how could the ads be effective in advancing consumer metrics as suggested by studies in marketing science? 

To understand this seeming contradiction, we phrase two null hypotheses in accordance to our previously defined research questions in the context of health-related search: 
\begin{itemize}
    \item \textbf{H1:}  when sponsored content is presented in the SERP, users will generally skip it and examine mostly organic results, preferring higher ranked results.
    \item \textbf{H2:} when users interact with a search engine to decide about the effectiveness of a medical treatment, the existence of sponsored content will not affect their decision. 
\end{itemize}

We will use the method developed in the next section to carefully study these research questions and verify our hypotheses. 



\section{Method and Experimental Setup}
We designed a series of user studies in which participants were required to evaluate the effectiveness of an intentionally selected natural remedy in treating a particular medical condition, only based on the entries in a SERP presented to them.

To this end, we created a dataset of SERPs with a wide variety of viewpoint biases (e.g., supporting or against the natural remedy's effectiveness), different configurations of the SERP (e.g., ranking positions of organic search results, with/without ads) and queries related to different remedies and health conditions. 
Insight regarding users' examination behaviour was collected by tracking users' clicks in their web browsers using JavaScript during the study. Insights regarding users' decision making and their search experience was collected explicitly via a series of survey questions by the end of the study.



\subsection{Dataset Preparation} \label{sec: data}
\subsubsection{Queries}
To conduct our study, we need a set of SERPs related to various queries about natural remedies' effectiveness for a  certain health condition. We chose the following three queries: \label{q:ginkgo} \textbf{Q1:} Ginkgo-biloba for treating tinnitus; \label{q:mlt} \textbf{Q2:} Melatonin for treating jetlag; \label{q:omega} \textbf{Q3:} Omega fatty acids  for treating ADHD.

Each of these three queries has a different ground-truth answer. According to their corresponding Cochrane reviews\footnote{Cochrane \cite{cipriani2011cochrane} is a charity organization whose mission is \say{\textit{to promote evidence-informed health decision-making by producing high-quality, relevant, accessible systematic reviews and other synthesized research evidence.}}
Cochrane only accepts conflict-free funding, so their reviewers can be considered as unbiased.}: Ginkgo biloba is \emph{not effective} in treating tinnitus~\cite{hilton2013ginkgo};  Melatonin is \emph{effective} in treating jet-lag~\cite{herxheimer2002melatonin}; and for omega-3 the answer is inconclusive and further study is required in order to reach a conclusion ~\cite{gillies2012polyunsaturated}. The detailed search result entries were manually retrieved by searching the above queries in well-known search engines. The query phrases entered to the search engines followed patterns that were designed to retrieve relevant results that contain a viewpoint regarding the effectiveness of the treatment for the condition discussed in the query. For example, ``\textit{is melatonin effective for jetlag}'', ``\textit{is melatonin ineffective for jetlag}'' etc.

\subsubsection{Advertisements}

A SERP can have one of three possible ad configurations: 
\begin{enumerate}
    \item Ads clean - no sponsored content at all; 
    \item Direct marketing ads;
    \item Indirect marketing ads. 
\end{enumerate}

Direct marketing ads are ads that offer the remedy in question for sale. This type of ads refer the users to an online shopping website where they can purchase the remedy. The shopping websites often do not directly discuss the effectiveness of the remedy, in particular not in the title or the snippet of the ads. For example, a direct marketing ad for omega fatty acids had the following title:``\textit{Omega-3 Fish Oil EPA DHA Triple Strength - 2,720 mg - 180...}'', and the following snippet:``\textit{Quality Vitamins \& Supplements Online At Bronson Vitamins. Quality Products Since 1960. 
Save an Additional 10\% To 20\% With Subscribe \& Save Auto Delivery Subscriptions.}''
We hypothesize that the existence of such products might indirectly suggest to users that the product is effective.
Direct marketing ads were retrieved from actual SERPs returned by the commercial search engines.

Indirect marketing ads are entries of sponsored content that contains textual content widely discussing and supporting the use of the remedy for the condition in question. Indirect marketing ads highlight the treatment's effectiveness already in their titles and snippets. For example, an indirect marketing ad for omega fatty acids had the following title: ``\textit{Omega 3s: The Ultimate (ADHD) Brain Food - ADDitude}'', and the following snippet: ``\textit{Supplementing with omega-3s eases hyperactivity. Analyzing data from 16 studies on ADHD and omega-3s, researchers at Oregon Health \& Science ...}'' Although these web pages do not focus on selling the product directly, all of them have links on their websites referring users to a shopping website. Indirect marketing ads were collected from organic search results that were annotated as having a positive viewpoint and containing direct marketing ads for the treatment in question as part of their web page. 
Notice that we excluded third party ad space such as Google AdSense but only considered static commercial content for the treatment in question
when classifying indirect marketing ads.

Our SERPs with sponsored content only included one ad (either direct or indirect marketing ad) placed at the top of the ranked list. Ads were marked by the bold text \textbf{Ad} to the left of their titles, as most commercial search engines would do. 
The manufactured SERPs resided on our experiment servers but the links to which the users were referred point to actual web pages found online, which we did not alter. 
Since our study's research questions focus on users' perception but not their general search experience, and given the high relevance of commercial search engines in presenting related ads to users, we opted to present only ads that were relevant to the search query at hand.

\subsubsection{Viewpoint Bias}\label{sec:viepoint}
The viewpoint of each entry was independently evaluated by two readers with necessary medical background  knowledge. Entries on which the readers disagreed were removed from the dataset. 
We categorize possible viewpoints into three classes: \emph{supporting (Y)}, \emph{rejecting (N)} and \emph{inconclusive (M)}, in order to coincide with the options later presented to our study participants. 


\begin{table}[t]
\caption{Bias level of the various SERP configurations, prior and posterior.}
\vspace{-3mm}
  \begin{tabular}{|c|c|c|c|c|}
    \hline
    Max Bias & B(Y) &B(M)&B(N) & Configurations\\

 \hline
            \multirow{4}{*}{Y} & & & & $y^am^an^a$\\\cline{5-5}
                                & 0.49-0.65 & 0.13-0.29 &0.13-0.29 &  $y^an^am^a$\\\cline{5-5}
             & (0.44-0.56)& (0.18-0.33)& (0.18-0.33)&$(ymn)^a$\\\cline{5-5}
             & & & &{$(ynm)^a$}\\\cline{5-5}             
 \hline
 \hline
            \multirow{4}{*}{M} & & & & $m^ay^an^a$\\\cline{5-5}
                                & 0.13-0.29 & 0.49-0.65 &0.13-0.29 &  $m^an^ay^a$\\\cline{5-5}
             & (0.18-0.33)& (0.44-0.56)& (0.18-0.33)&$(myn)^a$\\\cline{5-5}
             & & & &{$(mny)^a$}\\\cline{5-5}             
 \hline
\hline
            \multirow{4}{*}{N} & & & & $n^ay^am^a$\\\cline{5-5}
                                & 0.13-0.29 & 0.13-0.29 &0.49-0.65 &  $n^am^ay^a$\\\cline{5-5}
             & (0.18-0.33)& (0.18-0.33)& (0.44-0.56)&$(nym)^a$\\\cline{5-5}
             & & & &{$(nmy)^a$}\\\cline{5-5}             
 \hline
\end{tabular}
\label{tab:configs}
\vspace{-4mm}
\end{table}



All SERPs had the same number of entries for each of three possible viewpoints to avoid introducing other possible biases. However, they differed in the level of ranking bias induced by their actual rank positions.
We intentionally created two forms of ranking bias: block-wise ranking bias and interleaved ranking bias. In block-wise ranking bias, the viewpoints were divided into three blocks of consecutive result rankings about the same viewpoints. For example, the configuration $yyymmmnnn$ specifies a SERP with 9 results, where three supporting entries are placed in rank position 1-3, three inconclusive results in rank position 4-6, and three rejecting entries in rank position 7-9. 
In interleaved ranking bias, the viewpoints were interleaved with each other. For example, in the configuration $ymnymnymn$, three supporting entries are placed in rank positions \{1,4,7\}, three inconclusive results in rank positions \{2,5,8\} and three rejecting entries in rank positions \{3,6,9\}. We denote a sequence of $a$ consecutive viewpoints of value $v$ as $v^a$, and a sequence of interleaved viewpoints $v_iv_jv_k$ of length $a$ as $(v_iv_jv_k)^a$.
For example, the sequence $yyymmmnnn$ will be denoted as $y^3m^3n^3$ and the sequence $ymnymnymn$ will be denoted as $(ymn)^3$. 

The total a-priori ranking bias of a viewpoint in a SERP can  be viewed as the expected attention that the viewpoint receives.
Formally, the ranking bias that viewpoint $v$ expects to receive in a given result ranking $D$ of length $R$ is defined as:
\begin{equation} \label{eq:bias}
 B(D, v) = \frac{1}{Z}\sum_{i=1}^{R}P_c(i) \mathbbm{1}_{D[i]=v},
\end{equation}
where $Z$ is a normalization constant,  $\mathbbm{1}_{D[i]=v}$ suggests if the document placed at rank $i$ corresponds to viewpoint $v$ and $P_c(i)$ is the probability of the document to be examined by a user. Given no prior knowledge, we follow the position-based examination model \cite{craswell2008experimental} to estimate the probability by $P_c(i) \propto \frac{1}{i}$.

Table \ref{tab:configs} describes the possible SERP configurations grouped according to the viewpoint which receives the maximal ranking bias.
The first four, the middle four, and the last four are sequences where the maximal bias belongs to the supporting, inconclusive, and rejecting viewpoints respectively. 
The values in columns B(Y), B(M) and B(N) show the ranges of prior and posterior (in parenthesis) bias levels for the configuration associated with each maximal bias viewpoint, as was computed according to Eq \eqref{eq:bias}. 
The posterior bias values will be discussed in the experiment results section: they were calculated based on the actual click-through rate observed in our user study.

In our study the value $a$ was set to either 2 or 3 depending on the number of retrieved result entries for each viewpoint. For query Q1 (`Ginkgo-biloba for treating tinnitus'), we could not find three entries that expressed an inconclusive viewpoint, since its ground-truth is not effective; therefore in order to balance viewpoints, all SERPs for Q1 were limited to 6 results, i.e., $a=2$. For the other queries, $a$ was set to 3. 
For Q2 (`Melatonin for treating jetlag'), we could not find an indirect marketing ad entry. We assume this is because Melatonin can be easily purchased over the counter (at least in the US). Therefore our indirect marketing content experiments were only conducted for query Q1 and Q3.  

For each query there were 12 different bias configurations. For each query-configuration pair, three SERPs were randomly generated. To create the SERP instances with ads, a relevant ad of each type was added to each ad free SERP. In total, 288 SERPs were generated. These basic statistics about our study dataset are summarized in Table \ref{tab:doc_num}.


\begin{table}[t]
\caption{Statistics of the user study dataset.}
\vspace{-3mm}
  \begin{tabular}{|l|c|c|c|}
  \hline
  & $Q_1$ & $Q_2$ &$Q3$\\
  \hline
     a&2&3&3\\
    \hline
    \# of entries in a SERP without an ad & 6 &9&9\\
    \hline
    \# of entries in a SERP witho an ad & 7 &10&10\\
    \hline
    \# of SERP bias configurations & 12 & 12& 12\\
    \hline
    \# of SERP instance per configurations & 3& 3& 3\\
    \hline
    Total \# of unique SERPs  & 108& 72& 108\\
  \hline
\end{tabular}
\label{tab:doc_num}
\vspace{-2mm}
\end{table}

\subsection{Variables}
The dependant variable in this study is the participants' perceived effectiveness of a natural remedy for a given health condition. It is a categorical variable with three levels: effective, ineffective and inconclusive.  
The independent variables manipulated in our experiments were as follows:
\begin{enumerate}
    \item The discussed query (categorical; between-subjects). Each SERP is about a single query discussing the effectiveness of one treatment for a specific condition. 
    \item The sponsored content presented in the SERP (categorical; between-subjects).  A SERP could contain a direct marking ad, indirect marketing ad, or no ads at all.
    \item The viewpoint configuration presented in the SERP (categorical; between-subjects). 
    A SERP can take one of 12 possible viewpoint configurations listed in Table \ref{tab:configs}. 

\end{enumerate}

\subsection{User Study Procedure} \label{sec:survey}
In this section, we describe the procedure of the user study we conducted.  We used the Amazon Mechanical Turk (MTurk) platform to recruit participants for our study. 
All participants were from English speaking countries. We used MTurk's  external question format which directs participants to the website hosted by our survey web server.
Prior to the beginning of our study, we obtained an institutional review boards (IRB) approval from our institute. 

Upon entering the survey, a participant was requested to provide his/her demographic information, including age, gender, level and field of education. 
After filling in their personal information, the participant was directed to the instruction page. 
At this point, our server assigned a particular SERP instance to the participant. The SERPs were assigned to participants in a round robin manner to ensure a balanced number of responses for each experimental factor. 
Below is an example instruction page discussing the query `Ginkgo biloba for treating tinnitus':
\begin{quote}
Thanks for agreeing to participate in our research survey!
In the next page, you will be presented with a search result page for the query: \textit{is \textbf{Ginkgo biloba} an effective treatment for \textbf{tinnitus}}.
Your task is to determine the answer to this query, \underline{in the context} \underline{of the presented results only.}
Use the search results in the next page (click at least one of results) to find your answer. When you are ready, click the ``Answer Survey'' button to proceed.
\end{quote}

After reading the instructions, participants were asked if they had any previous knowledge regarding the topic of the study. The available options were: 1) yes I do, 2) not a lot, and 3) none. 
We did not want to collect responses that were resulted from the participants' prior knowledge or bias, but we also did not want them to lie in order to participate for payment. Therefore, we did not use this question to filter any participants; and for those who declared having prior knowledge regarding the query, their responses were excluded from the entire analysis of our study (they were still paid for their participation). This was done as part of our study design. 

After answering this question, the participants were referred to the assigned SERP to start the task. 
They were instructed to browse as many links as they needed to answer the question; and they could not progress to the next stage unless they had clicked at least one link. This was to ensure the participants understood that they had to interact with the SERP. 

Once the participants pressed the ``Answer Survey'' button that was located next to the search box, they were then directed to the question form, where they were required to enter their conclusion regarding the effectiveness of the treatment discussed in the query. The possible answers were: 
\begin{itemize}
    \item No - according to the information I read, it is not effective.
    \item Yes - according to the information I read, it is effective.
    \item Maybe - according to the information I read, it is inconclusive whether or not it is effective.
    \item Not sure - there was not enough information for me to reach a conclusion.
\end{itemize}

In this form, we also included 3 attention questions to filter out careless workers. Participants were required to  1) answer  a multiple choice question about the nature of the remedy that the query was intended for, 2) answer a multiple choice question about the nature of the medical condition that the query concerned, and 3) provide a reason or piece of evidence for their conclusion regarding the remedies' effectiveness. 

The attention question about the nature of the remedy  was phrased simply as `What is REMEDY'. The available answers for it were:
\begin{itemize}
    \item A hormone (True for melatonin)
    \item A nutrient found mostly in fish, nuts and seeds (True for omega fatty acids)
    \item A tree (True for Ginkgo biloba)
\end{itemize}

The attention question about the nature of the medical condition was phrased simply as `What is MEDICAL CONDITION'. The available answers for it were:
\begin{itemize}
    \item A neurological disorder (True for ADHD)
    \item Ringing noise in one or both ears (True for tinnitus)
    \item A sleep problem (True for jet-lag)
\end{itemize}

Participants who did not answer the filter questions correctly or did not provide a coherent reason for their choice, were removed from the analysis.
Participants were also provided with the opportunity to add general comments about the study itself. By the end of our study, we received very positive feedback from the participants, indicating that the survey instructions were clear and the task  was easy to follow. Several participants noted that they liked the survey and it was interesting to them. 

Once participants submitted their answer about the treatments' effectiveness and attention questions, they could not go back and alter them. 
If the SERP presented to the participants included ads, they were presented with two additional questions about the influence of the ad to their overall search experience and to their decision about the treatment's effectiveness. Participants were also allowed to enter any general comments concerning the ad. 
The first question 
was phrased as:
``The search result page you viewed contained sponsored content (Ads). How much did the ad affect your search experience?''
The responses to this question were measured by a 5-point Likert scale which ranged from -2 to +2, with negative values indicating the ads worsened the search experience and positive values indicating it improved it.
The second question was 
phrased as: ``How much did the ad affect your decision regarding the treatment's effectiveness?'' We again used a 5-point Likert scale with negative values indicating the ads added doubts to participants' decisions and positive values indicating ads enhanced their belief in their decision. 

Once the participants answered all of the survey's questions, they would receive a verification code to receive their payment from the MTurk website. Since our study is between subject, each participant was exposed to only one SERP. Once a participant completed the task, they were not offered to take the survey again. 
On this page, they were also presented with a highlighted warning message noting that they were participating in an experiment and should not make any real conclusions as to the true effectiveness of the remedy discussed in the query.

Participants were paid at a basis rate of \$1 for participation. In addition, they were informed that they would receive a \$1 bonus if their answers were qualified based on their logged search behaviors. By our design, to receive the bonus they had to answer our post survey attention questions correctly and spend at least two minutes on the task itself. 
The payment scheme was derived from a preliminary experiment run in our lab by volunteers that were not MTurk workers, in order to estimate the time it takes to complete the survey task. We added additional time to account for slower working participants, in order to make sure we comply with an hourly payment of at least the US minimum wage. A later analysis of the average time it took the actual participants to complete the task reveled that this was indeed the case.

\begin{figure*}[t]
\centering
\begin{subfigure}{0.33\textwidth}
\centering
  \includegraphics[scale=0.33,keepaspectratio]{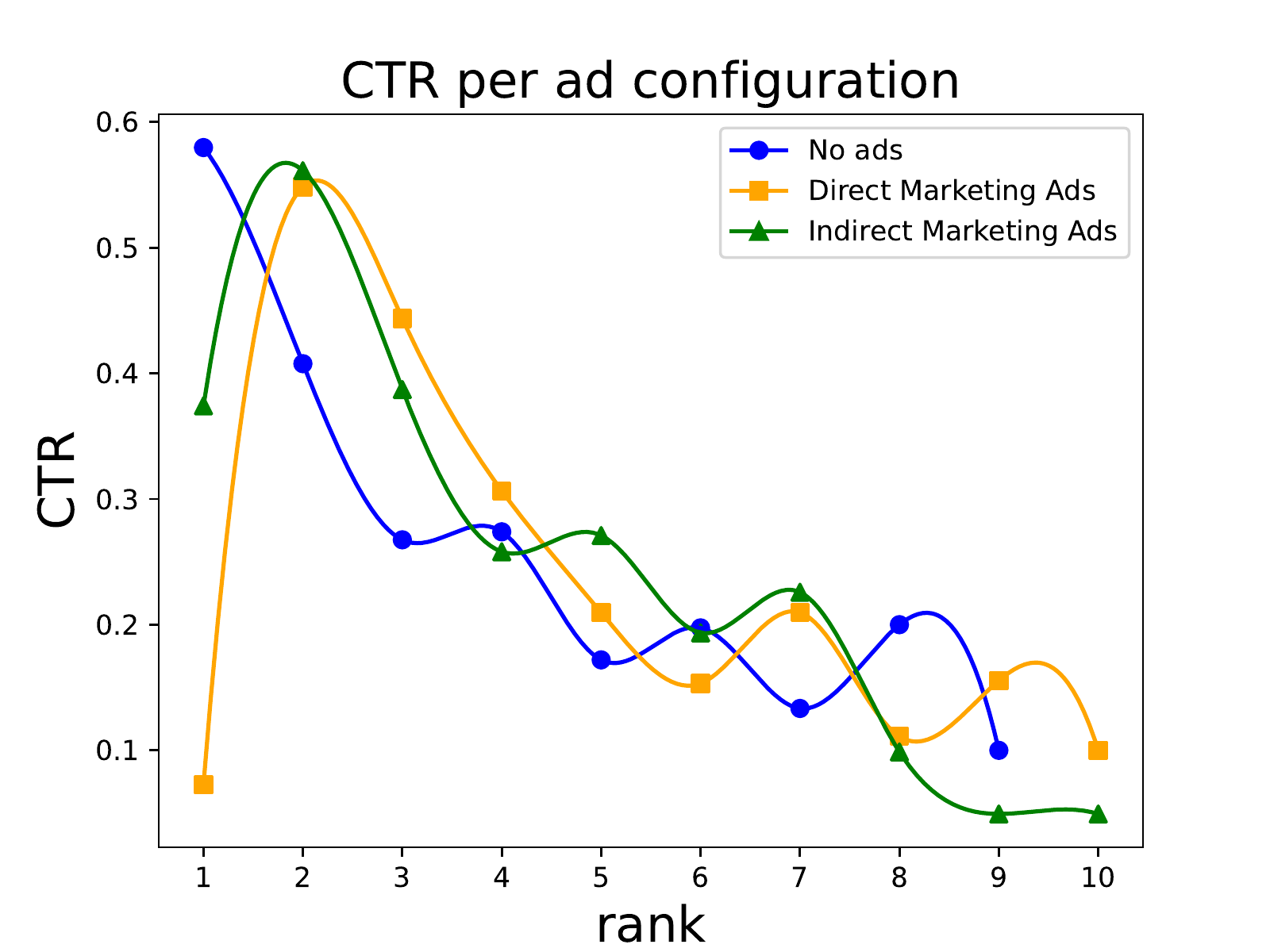}
  \caption{All marketing Ads.}
  \label{fig:ctr}
\end{subfigure}
\begin{subfigure}{0.33\textwidth}
\centering 
 \includegraphics[scale=0.33,keepaspectratio]{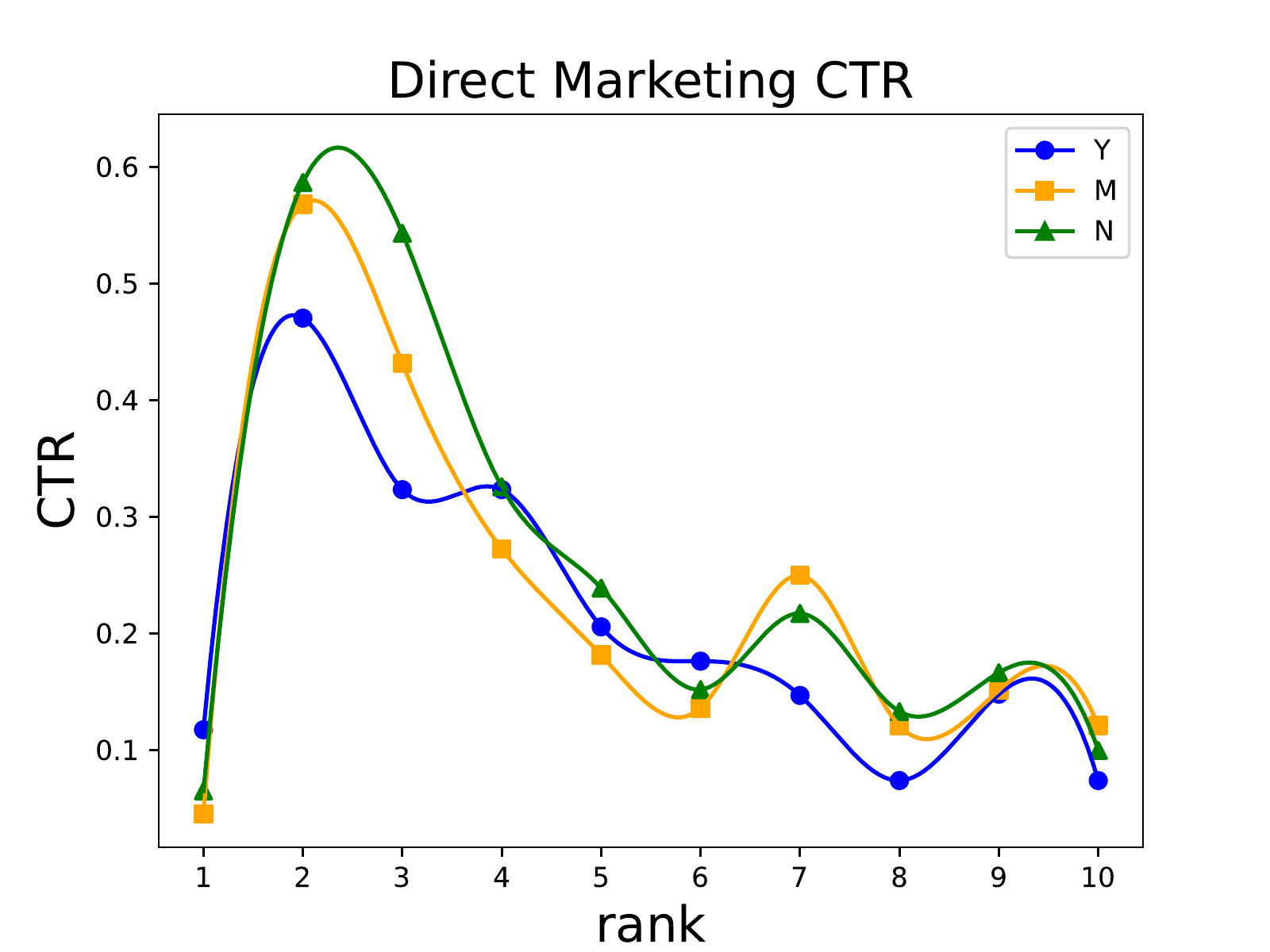}
  \caption{Direct marketing Ads only.}
  \label{fig:direct}
\end{subfigure}
\begin{subfigure}{0.33\textwidth}
  \centering
\includegraphics[scale=0.33,keepaspectratio]{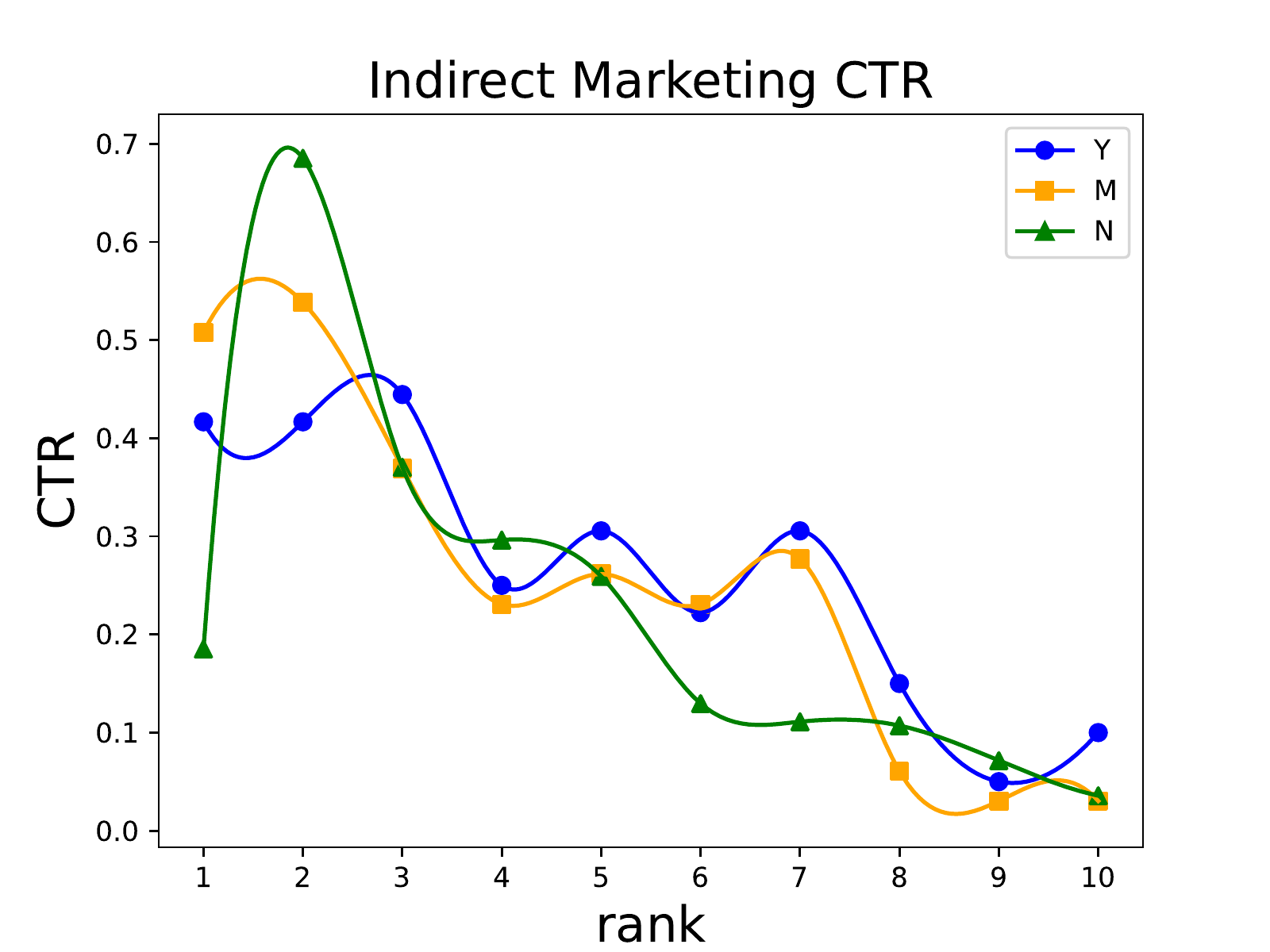}
  \caption{Indirect Marketing Ads only.}
  \label{fig:spon}
\end{subfigure}%
\vspace{-2mm}
\caption{CTR comparison according to maximal bias.}
\label{fig:ctr_ads}
\vspace{-2mm}
\end{figure*}

\begin{table}[t]
\caption{Demographic information of study participants.}
\vspace{-1mm}
\begin{tabular}{ |l|c|c| } 
 \hline
 Age range & 20-74 \\
 \hline
 Average age  & 39 (sd=11) \\
 \hline
 Men & 63\% \\
 \hline
 Women & 37\% \\
 \hline
 With bachelor's degree and above & 61\%\\
 \hline
 No higher education & 39\% \\
 \hline
\end{tabular}
\label{tab:demographics}
\vspace{-5mm}
\end{table}

\section{Experiment results}
In this section, we will answer our previously defined research questions based on the observations collected from our user study.
Our research questions focus on two important aspects: sponsored content's impact on users' examination behaviour and that on users' decision-making. Section \ref{sec:CTR} will discuss the research question related to users' examination behaviour, and Section \ref{sec:dec} will discuss the research question related to users' decision making. We will also provide summary statistics related to users' self-perception about the effect of sponsored content on their search experience and decision making in Section \ref{sec:exp}.

We filtered out responses from users who did not answer our attention question correctly (81 responses) and responses from users who did not provide any reason supporting their responses (63 responses), which is explicitly required in our instructions. Example of bad reasons provided by users include empty phrase or single word sentences, vague answers such as: ``it provides correct information''. Phrases only describing the treatments or the condition without referring to the effectiveness of the treatment were also removed. 

After filtering, our user study produced 436 valid responses from 436 different participants. Each participant was assigned to a single SERP (i.e., ad configuring-bias level-query condition combination). 
Table \ref{tab:demographics} summarizes the demographic information of the participants, and Table \ref{tab:ads_dist} summarizes the number of responses received for the different settings of each independent variable. In average, each different ad configuration-bias-query triple received 18 observations (std=6).
Participants spent an average of 4.7 minutes on a given SERP (std=3.17) and clicked 2.3 links  (std=1.36) in our study. In total, 992 links were entered by all users, of which 925 (93 \%) were organic results and 67 (7\%) were links to sponsored content. 

\begin{table}[t]
\caption{Levels and descriptive statistics of observations per independent variable}
\vspace{-1mm}
\begin{tabular}{ |c|c|c| } 
\hline
\thead{Independent \\ Variable} & \thead{Levels} & \thead{\#Responses } \\
 \hline
\multirow{3}{*}{
\makecell{Ad \\ Configuration} }& No ads & 157\\
\cline{2-3}
& Direct Marketing &124\\
\cline{2-3}
&Indirect Marketing&155  \\
  \hline
  \hline
\multirow{3}{*}{Query} & Ginkgo-Biloba for  tinnitus&145\\
\cline{2-3}
&Melatonin for jetlag&105\\
\cline{2-3}
&Omega fatty  for treating ADHD & 186\\ 
 \hline
\hline
\multirow{3}{*}{SERP Bias} &Y&125\\
\cline{2-3}
&M&159\\
\cline{2-3}
&N&152\\
\hline
\end{tabular}
\label{tab:ads_dist}
\vspace{-6mm}
\end{table}

\subsection{Impact on Examination 
Behaviour}\label{sec:CTR}

To understand the difference in examination behaviour between the different ad configurations, we examine click-through rate (CTR) of each rank position in the SERP, for the various ads and bias configurations. CTR is defined as the number of times a link was clicked, divided by the number of times it was visible to the users.

Let $r_i$ denote rank position $i$ in a given SERP and $o_i$ denote the rank of $i$th organic result \footnote{An organic result is a search result that is not sponsored} in a SERP. In a SERP without ads, the general documents' ranking and organic results' ranking are the same, therefore $r_i$ equals to $o_i$. When ads are present however, $r_i$ equals to $o_{i-1}$, i.e., the second ranked result is the first organic result and so forth. And lastly, let $CTR(r_i)$ and $CTR(o_i)$ denote the CTR of $i$th ranked document and the $i$th organic result in accordance. 

Figure \ref{fig:ctr} presents CTR over rank positions, for each of the three possible ad configurations. At a high level view of the results, it is evident that the examination behaviour of users is different when ads are present. It is easy to notice that direct marketing ads are generally ignored by users and indirect marketing ads receive significantly less clicks than organic results at the same rank position. 


As can be seen from the figure,
when ads are not presented (i.e., the first ranked position is organic result), the first position receives the most attention from users, with a CTR at around 0.58. This is in comparison to a CTR of 0.37 when the first ranked result is an indirect marketing ad and only 0.07 when the first ranked result is a  direct marketing ad.
A one-way ANOVA test comparing $CTR(r_1)$ among all three ad configurations revealed that this difference is statistically significant (F(2,433)=46, p<0.01). A further post-hoc Tukey’s HSD test for multiple comparisons revealed that the mean value of each pair of configurations was significant as well, meaning that there is a significant difference not only between $CTR(r_1)$ in SERPs with no ads compared to SERPs with ads, but also between the two different ad configurations, with p < 0.01 for all comparisons. 

In the second ranked position, we see that for ad-free SERPs the CTR declines from 0.58 to 0.4. In contrast, the exhibited examination behaviour when ads were presented shows an increased CTR on the second position from 0.37 to 0.56 for SERPs with indirect marketing ads and from 0.07 to 0.55 for SERPs with direct marketing ads, values similar to that of $r_1$ in the no ad configuration. This suggests the first organic result receives similar attention in all ad configurations. A one-way ANOVA test of $CTR(r_2)$ between the three ad configurations again showed a statistically significant difference (F(2,433)=46, p<0.02). This time the post-hoc Tukey’s HSD test revealed a significant difference only between the `No Ads'  and each ad configuration, but not between the two ad configurations.
This clearly demonstrates that, despite the ad's position advantage, participants showed a clear preference towards organic results; or in other words, a strong tendency to skip the ads.

To better understand the effect of the different ad configurations on users' examination behaviour,
we further decompose the CTR of the two ad configurations according to the maximal a priori bias level in the SERP, as detailed in Table \ref{tab:configs}. 
Figures \ref{fig:direct} and \ref{fig:spon} show the CTR distribution by maximal bias level for SERPs with direct and indirect marketing ads.

As can be observed in Figure \ref{fig:direct}, for SERPs with direct marketing ads, a similar examination behaviour is exhibited across all bias levels: The CTR of the first position is generally very low, while the CTR of the second position receives significantly higher attention. 


But surprisingly, this is not the case in SERPs with indirect marketing ads. As shown in Figure \ref{fig:spon}, participants' behaviour is not uniform across all bias levels. The curve for the negative bias configuration (i.e., `N') behaves quite similarly to that of the curves presented in Figure  \ref{fig:direct}, with a very low CTR for $r_1$ (0.18) that jumps to a very high CTR (0.68) for $r_2 / o_1$, and declines to 0.37 for $r_3/o_2$.
This suggests that when the bias of the SERP is negative the ads are generally ignored by the users. When we inspected the other two bias configurations, however, we encountered different observations. In these configurations, we do not see the behaviour exhibited in the direct marketing ad setting. 
And on the other hand, the behaviour is not similar to that exhibited by the no ad configuration either, where $CTR(r_1)$ is significantly higher than $CTR(r_2)$ and on wards. 
In SERPs with positive bias (i.e., `Y'), we see that the CTRs for the first three rank positions are similar to each other at around 0.4. The significant decline only begins at $r_4/o_3$, which drops to 0.25. In SERPs with inconclusive bias (i.e., `M'), we see a similar CTR of the two highest rank positions of 0.51 and 0.54, followed by a decrease to 0.37 for $r_3/o_2$.

We believe the reason for these observations stems from the level of contrast between viewpoints of the organic and sponsored content. Bear in mind that indirect marketing ads are essentially web pages that present content supporting the effectiveness of the treatment. This supporting opinion is evident not only by reading the content but also from the title and snippet of the presented results in the SERP.
When the SERP's bias level is towards the negative viewpoint, the opinion in the ad is strongly contradicted by the organic results, i.e., the contrast between the ad's viewpoint and the bias of the organic result is high.
The participants can notice this fact, skip the ad, and proceed to click on the organic results.
However, when the contrast is
not high, participants are more prone to inspect the content of the ad. 
We hypothesize that because the indirect marking ads typically have rich content, e.g., its title and abstract often contain professional or medical terms, when the contrast level is low, the users tend to reduce their negativity towards sponsored content and are more willing to open the link for further investigation. Please note in our setting one MTurk worker can only encounter one SERP configuration and thus they were not aware of the different biases manufactured in our study. As we will present in Section \ref{sec:exp}, interestingly, although almost all our participants reported they were not affected by the presence of sponsored content, their examination behaviors already differed conditioned on what types of sponsored content were presented.

Our observations lead to mostly reject hypothesis H1: Although in some cases, users do skip ads and prefer higher ranked organic results, there is still a non-negligible portion of cases where the users inspect the ads. 




\subsection{Impact on Decision-Making}\label{sec:dec}

\begin{table}
\caption{Participants' responses absolute and percentages values, grouped by ads-bias configurations.}
\vspace{-1mm}
 \begin{tabular}{|c|c|c|c|}
\cline{2-4}
\multicolumn{1}{c|}{}  & \textbf{Yes} & \textbf{Maybe} &\textbf{No}\\
\hline
\textbf{Y}&28(0.52)&14(0.26)&12(0.22)\\
\hline
\textbf{DY}&15(0.45)&14(0.42)&4(0.12)\\
\hline
\textbf{IY}&14(0.41)&15(0.44)&5(0.15)\\
\hline
\hline
\textbf{M}&12(0.24)&25(0.50)&13(0.26)\\
\hline
\textbf{DM}&11(0.26)&20(0.47)&12(0.28)\\
\hline
\textbf{IM}&25(0.39)&27(0.42)&12(0.19)\\
\hline
\hline
\textbf{N}&7(0.14)&22(0.44)&21(0.42)\\
\hline
\textbf{DN}&10(0.22)&16(0.35)&20(0.43)\\
\hline
\textbf{IN}&9(0.17)&14(0.27)&29(0.56)\\ 
\hline
\end{tabular}
\label{tab:user_conf}
\vspace{-2mm}
\end{table}

\begin{table}
\caption{T-Tests comparisons between the different answers for all ad-bias configurations.  \checkmark\ \checkmark stands for $p< 0.01$,  \checkmark  for $p < 0.05$, and $\times$ for $p \geq 0.05$.}
\vspace{-1mm}
 \begin{tabular}{|c|c|c|c|}
\cline{2-4}
\multicolumn{1}{c|}{}&\thead{Yes-\\Maybe}&\thead{Yes-\\No}&\thead{Maybe-\\No}\\
\hline
\textbf{Y}&\checkmark \checkmark&\checkmark \checkmark&$\times$\\
\hline
\textbf{DY}&$\times$&\checkmark \checkmark&\checkmark \checkmark\\
\hline
\textbf{IY}&$\times$&\checkmark \checkmark&\checkmark \checkmark\\
\hline
\hline
\textbf{M}&\checkmark \checkmark&$\times$&\checkmark \checkmark\\
\hline
\textbf{DM}&\checkmark \checkmark&$\times$&\checkmark \checkmark\\
\hline
\textbf{IM}&$\times$&\checkmark \checkmark&\checkmark \checkmark\\
\hline
\hline
\textbf{N}&\checkmark \checkmark&\checkmark \checkmark&$\times$\\
\hline
\textbf{DN}&\checkmark &\checkmark \checkmark&\checkmark \\
\hline
\textbf{IN}&\checkmark &\checkmark \checkmark&\checkmark \checkmark\\
\hline

\end{tabular}
\label{tab:ttest}
\vspace{-3mm}
\end{table}

We now focus on the research question about users' decision-making process.

Table \ref{tab:user_conf} shows the distribution of participants' responses grouped by ad configuration and the maximal level of posterior bias computed by Eq \eqref{eq:bias}.
The posterior bias level is computed in a similar way as the priori bias level, 
but by using the actual CTR observed in our study data: $P_c(i) = CTR(r_i)$. As can be seen in Table \ref{tab:configs}, while the absolute values of the posterior and priori bias levels are slightly different, the order among viewpoint bias remained the same. This means that 
our participants behaved as expected according to our existing knowledge on position bias \cite{craswell2008experimental}. 

Each row in Table \ref{tab:user_conf} corresponds to an ad configuration and a bias level. The rows noted by a single letter (`Y', `M', `N') describe the response distribution across different bias levels of ad-free SERPs, and the rows marked by two letters describe users' response distribution across different bias levels for SERPs with either direct (D) or indirect (I) marketing ads. 
The columns represent the number of responses  along with the percentage of responses in parenthesis:  ``Yes'' for a positive response, ``Maybe'' for an inconclusive response, and ``No'' for a negative response. Out of 436 responses, only 10 were `No Sure'. For clarity of presentation, they were therefore discarded from the analysis.
Table \ref{tab:ttest} summarizes the t-test results of the portion of results each answer received for the different ad-bias configurations. 

From the results we can see that in the no ads configurations, there is a correlation between the maximal bias level and the most popular response, which aligns with previous work on position bias.
This is very evident for the ``Y'' and ``M'' bias levels and less evident for the ``N'' bias level. We can also see that for each bias configuration a different user behaviour pattern is exhibited in at least one of the ad configurations.

For indirect marketing ads, we can observe ads' influence in all bias levels, with the direction of the influence depending on the SERP`s bias level. When the viewpoint bias of the SERP is either positive or inconclusive (i.e. rows ``IY'' and ``IM''), we observe a decrease in the portion of negative responses in both configurations. For the ``IY'' configuration, we then observe an increase in the portion of ``Maybe'' responses; and for the ``IM'' configuration,  an increase of the ``Yes'' responses. 
We also observe that in both configurations the ``Maybe'' and ``Yes'' responses receive a similar portion of the answers. These observations coincide with the CTR data for these configurations. For the ``IM'' configuration, we observed that the first 2 positions received similar attention from partipants. These positions represent an ad and an inconclusive viewpoint organic result. For the ``IY'' configuration, we observed that the first 3 positions received similar attention. These positions represent an ad, a positive viewpoint organic result, and another organic result that can belong to either viewpoint. In these configurations, when indirect marketing ads are present, users were exposed to a wider viewpoint distribution in comparison to the corresponding no ads configuration, which in turn affected their decisions. 

Direct marketing ads are rarely inspected according to our CTR data. Nevertheless they still affected users' decisions, although in a more subtle manner than indirect marketing ads. The effect, when exists, is always towards the positive side. 
For the ``DY'' configuration we observe a shift from ``No'' to ``Maybe''.
We observe no significant difference for the ``DM'' configuration; and for the ``DN'' configuration we see a shift from ``Maybe'' to ``Yes''. 

We hypothesize that the difference between partipants' reaction to the two ad types is due to the difference between the manner in which these two  types support the treatment`s effectiveness. Direct marketing ads offer the treatment for sale but do not explicitly support its effectiveness. Indirect marketing ads explicitly support the treatment's effectiveness by eagerly announcing its helpful qualities already in the title and snippet of the ad. Indirect marketing, therefore, elicits stronger sentiment, either positive or negative depending on the organic results presented in the SERP.

\begin{figure}[t]
\centering
\includegraphics[width=\linewidth]{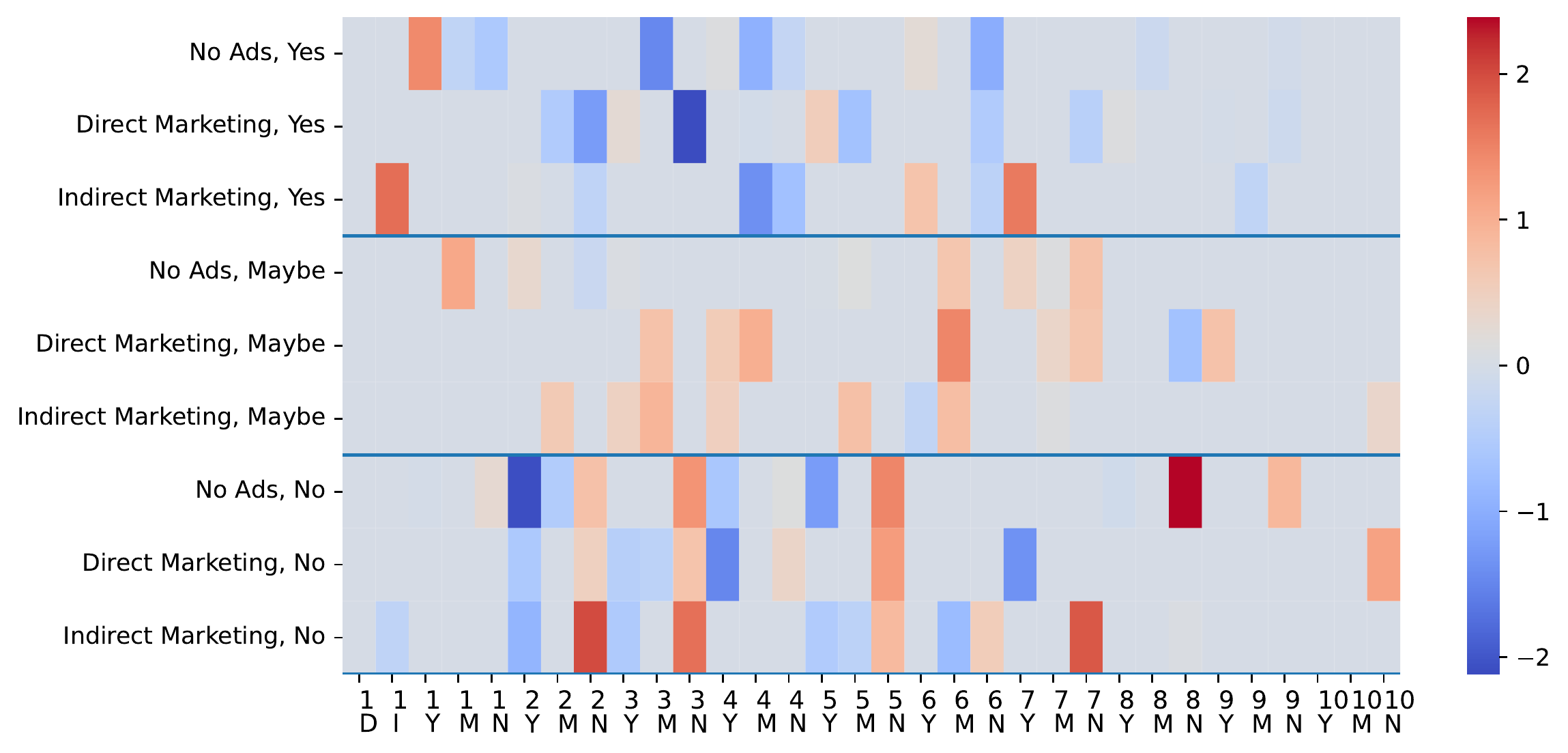}
\vspace{-3mm}
\caption{HeatMap for logistic regression learnt feature weights.}
\label{fig:HeatMap}
\vspace{-3mm}
\end{figure}
To better understand the connection between ads and users' decisions, we also examine the effect of ads on the correlation between users' examination behaviour and their decisions. To achieve so, we trained a set of multi-class logistic regression classifiers and analyzed their learnt feature weights. These classifiers were fitted to predict a user's response given the SERP presented to her based on her observed clicks.
A model was trained for each of the three possible ad configurations independently. 
Binary features in the form of $f_i^l=\{0,1\}$ were constructed to encode users' clicks, where $i$ denotes the position of a document and $l$ denotes the viewpoint label (`Y', `M' or `N') or the ad configuration (`D' or `I') of that document. $f^i_l=1$ if the document at position $i$ with label $l$ was clicked, and 0 otherwise. For example $f^1_D=1$ means that in one experiment the user clicked on the direct marketing ad at position 1.
Figure \ref{fig:HeatMap} shows a heat-map of the learnt weights. We used  $\ell_1$ regularization for feature selection purposes, to help the classifiers learn the most significant features. The X axis denotes the features $f^i_l$ and the Y axis denotes the ad configuration and the user's response class (comma separated) under which the model was trained.

Inspecting the heat map, we can observe a general correlation between the viewpoint of users' inspected content and their decision, validating our experimental methodology.
For the models predicting ``Yes'' and ``No'' answers, only correlating viewpoints (i.e  features of the form $f^1_Y$ for ``Yes'' and $f^1_N$ for ``No'' ) received a positive weight, in all ad configurations. For the models predicting ``Maybe'' we observe positive influence of entries from all three viewpoints, however no influence from either ad type. 
We can also observe that consuming the content of indirect marketing ads influences users to select a positive response, with ($f^1_I$) receiving a positive weight in the model ``Indirect Marketing, Yes'' and a negative weight in the model  ``Indirect Marketing, No''. This suggests that when consumed, indirect marketing ads introduce positive bias to the decision making process. We do not observe any correlation between clicking on direct marketing ads and the various responses, as expected given their low CTR. Bear in mind that the regression only inspected the effect of positions that were clicked and cannot infer the effect of consuming the ads' title and snippet content without clicking on the link.

We are now ready to answer our research questions whether sponsored content affects users' decision making. And our answer is affirmative. We therefore also reject H2 which hypothesised that ads do not affect users decision making. 
As our results indicate, sponsored content can introduce either positive or negative bias, depending on the type of the ad as well as on the bias presented by the organic results. 

\subsection{Impact on Search Experience}\label{sec:exp}

\begin{figure}[t]
\centering
\includegraphics[width=0.7\linewidth,scale=0.2,keepaspectratio]{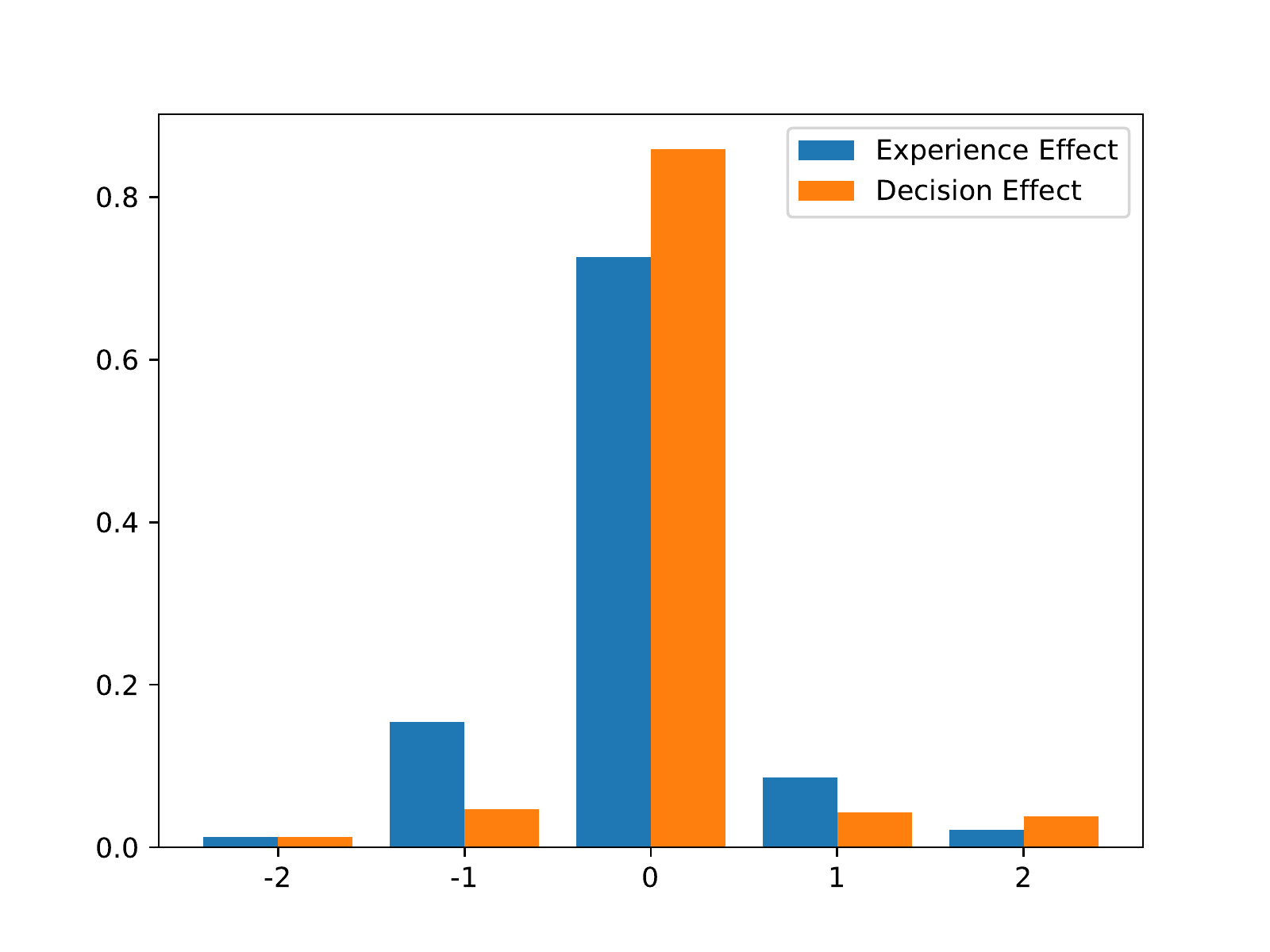}
\vspace{-6mm}
\caption{Ads effect on users' experience.}
\label{fig:user_reports}
\vspace{-6mm}
\end{figure}

Participants whose SERP contained ads were requested to rate the ads' effect on their search experience and their decision-making process. The distributions of participants' responses for both questions are reported in Figure \ref{fig:user_reports}. 
The vast majority of participants reported that the ad did not affect either their search experience or their decision-making. There was no significant difference between the two ads configurations.

69 participants provided comments about the effect of the ad on their experience. Most of the participants (58) comments were neutral, noting that they did not mind or notice the ad in the experiments, and they said they tended to ignore ads in general.
For example: 
``\textit{I ignored the ad that I saw and skipped down to the first non-ad link.}'',
 and ``\textit{Honestly, I don't look at ads. I never open them when I'm researching.}'' 
Some participants expressed dissatisfaction in their comments. For example:
``\textit{Intrusive and annoying.  The world is one big ad.}''. 
And some admitted the negative bias that ads invoke in them, for example: ``\textit{I notice them, but automatically disregard them as unreliable.}'',  ``\textit{I usually ignore ads, I don't trust them.}'', and `` \textit{I generally skip over the ads when looking for actual information.  I'm not going to go with a paid ad for research purposes.}''.
Only one participant had positive comments about the ads noting that they were informative and relevant to the search.  

To summarize, the participants' responses and comments indicate that the majority of them perceived themselves as unaffected by ads. However, the results presented in the previous sections suggested that oftentimes this is not the case, as both their examination and decision-making are affected by the sponsored content presented in the SERPs. 

 

\section{Discussion and conclusions}
In this work, we conducted a series of user studies to examine the effect of sponsored content on users' examination behaviour and decision-making process in medical-related web searches. Our participants were presented with SERPs with different levels of biases towards distinct viewpoints and three different types of sponsored content configurations: no ads, direct marketing ads and indirect marketing ads. 
We found that users reacted differently to direct marketing ads when compared to indirect marketing ads, depending on the bias in the organic content. 
As supported by previous results, we found that the ranking bias in the SERP significantly affected users' decision-making. 
Direct marketing ads were generally skipped by the users who preferred to click on organic results. However, their mere existence still affected users decisions. Indicating that although the ads were skipped, they were not ignored.  
When users were presented with indirect marketing ads, their reactions depended on the contrast between the positive viewpoint presented by the ad and the viewpoint bias presented in the SERP.
A strong contrast caused participants to skip the ads and introduced additional negative bias.  In weak and no-contrast settings, however, participants were more inclined to inspect the ads, which in turn introduced positive biased to their decisions. 
Moreover, although the majority of our study participants believed their search behaviors were not affected by the sponsored content, the statistics showed the opposite, suggesting the influence is even unconsciously.

Users in this age of internet are extremely aware of the existence of sponsored content and the interests they represent. They dislike ads but perceive themselves as not affected by them. The results of our study showed that when sponsored content is sophisticated enough it can neutralize this negative bias. However, the organic results' viewpoint still take precedent over the sponsored content.
The accuracy of search results in the health domain is therefore crucial in mitigating bias and misinformation. 





\bibliographystyle{ACM-Reference-Format}
\bibliography{main}


\begin{thebibliography}{40}


\ifx \showCODEN    \undefined \def \showCODEN     #1{\unskip}     \fi
\ifx \showDOI      \undefined \def \showDOI       #1{#1}\fi
\ifx \showISBNx    \undefined \def \showISBNx     #1{\unskip}     \fi
\ifx \showISBNxiii \undefined \def \showISBNxiii  #1{\unskip}     \fi
\ifx \showISSN     \undefined \def \showISSN      #1{\unskip}     \fi
\ifx \showLCCN     \undefined \def \showLCCN      #1{\unskip}     \fi
\ifx \shownote     \undefined \def \shownote      #1{#1}          \fi
\ifx \showarticletitle \undefined \def \showarticletitle #1{#1}   \fi
\ifx \showURL      \undefined \def \showURL       {\relax}        \fi
\providecommand\bibfield[2]{#2}
\providecommand\bibinfo[2]{#2}
\providecommand\natexlab[1]{#1}
\providecommand\showeprint[2][]{arXiv:#2}

\bibitem[PEW(2012)]%
        {PEW_trust}
 \bibinfo{year}{2012}\natexlab{}.
\newblock \bibinfo{title}{SEARCH ENGINE USE 2012}.
\newblock
  \bibinfo{howpublished}{\url{https://www.pewresearch.org/internet/2012/03/09/main-findings-11/}}.
\newblock
\urldef\tempurl%
\url{https://www.pewresearch.org/internet/2012/03/09/main-findings-11/}
\showURL{%
\tempurl}


\bibitem[PEW(2019)]%
        {PEW}
 \bibinfo{year}{2019}\natexlab{}.
\newblock \bibinfo{title}{{Majority of Adults Look Online for Health
  Information | PEW Research Center}}.
\newblock
  \bibinfo{howpublished}{\url{https://www.pewresearch.org/fact-tank/2013/02/01/majority-of-adults-look-online-for-health-information/}}.
\newblock
\urldef\tempurl%
\url{https://www.pewresearch.org/fact-tank/2013/02/01/majority-of-adults-look-online-for-health-information/}
\showURL{%
\tempurl}


\bibitem[Agarwal et~al\mbox{.}(2015)]%
        {agarwal2015organic}
\bibfield{author}{\bibinfo{person}{Ashish Agarwal}, \bibinfo{person}{Kartik
  Hosanagar}, {and} \bibinfo{person}{Michael~D Smith}.}
  \bibinfo{year}{2015}\natexlab{}.
\newblock \showarticletitle{Do organic results help or hurt sponsored search
  performance?}
\newblock \bibinfo{journal}{\emph{Information Systems Research}}
  \bibinfo{volume}{26}, \bibinfo{number}{4} (\bibinfo{year}{2015}),
  \bibinfo{pages}{695--713}.
\newblock


\bibitem[Alanazi et~al\mbox{.}(2020)]%
        {alanazi2020impact}
\bibfield{author}{\bibinfo{person}{Afrah~Olayan Alanazi}, \bibinfo{person}{Mark
  Sanderson}, \bibinfo{person}{Zhifeng Bao}, {and} \bibinfo{person}{Jaewon
  Kim}.} \bibinfo{year}{2020}\natexlab{}.
\newblock \showarticletitle{The impact of ad quality and position on Mobile
  SERPs}. In \bibinfo{booktitle}{\emph{Proceedings of the 2020 Conference on
  Human Information Interaction and Retrieval}}. \bibinfo{pages}{318--322}.
\newblock


\bibitem[Allam et~al\mbox{.}(2014)]%
        {allam2014impact}
\bibfield{author}{\bibinfo{person}{Ahmed Allam},
  \bibinfo{person}{Peter~Johannes Schulz}, {and} \bibinfo{person}{Kent
  Nakamoto}.} \bibinfo{year}{2014}\natexlab{}.
\newblock \showarticletitle{The impact of search engine selection and sorting
  criteria on vaccination beliefs and attitudes: two experiments manipulating
  Google output}.
\newblock \bibinfo{journal}{\emph{Journal of medical internet research}}
  \bibinfo{volume}{16}, \bibinfo{number}{4} (\bibinfo{year}{2014}),
  \bibinfo{pages}{e100}.
\newblock


\bibitem[Burke et~al\mbox{.}(2005)]%
        {burke2005high}
\bibfield{author}{\bibinfo{person}{Moira Burke}, \bibinfo{person}{Anthony
  Hornof}, \bibinfo{person}{Erik Nilsen}, {and} \bibinfo{person}{Nicholas
  Gorman}.} \bibinfo{year}{2005}\natexlab{}.
\newblock \showarticletitle{High-cost banner blindness: Ads increase perceived
  workload, hinder visual search, and are forgotten}.
\newblock \bibinfo{journal}{\emph{ACM Transactions on Computer-Human
  Interaction (TOCHI)}} \bibinfo{volume}{12}, \bibinfo{number}{4}
  (\bibinfo{year}{2005}), \bibinfo{pages}{423--445}.
\newblock


\bibitem[Buscher et~al\mbox{.}(2010)]%
        {buscher2010good}
\bibfield{author}{\bibinfo{person}{Georg Buscher}, \bibinfo{person}{Susan~T
  Dumais}, {and} \bibinfo{person}{Edward Cutrell}.}
  \bibinfo{year}{2010}\natexlab{}.
\newblock \showarticletitle{The good, the bad, and the random: an eye-tracking
  study of ad quality in web search}. In \bibinfo{booktitle}{\emph{Proceedings
  of the 33rd international ACM SIGIR conference on Research and development in
  information retrieval}}. \bibinfo{pages}{42--49}.
\newblock


\bibitem[Cipriani et~al\mbox{.}(2011)]%
        {cipriani2011cochrane}
\bibfield{author}{\bibinfo{person}{A Cipriani}, \bibinfo{person}{TA Furukawa},
  {and} \bibinfo{person}{C Barbui}.} \bibinfo{year}{2011}\natexlab{}.
\newblock \showarticletitle{What is a Cochrane review?}
\newblock \bibinfo{journal}{\emph{Epidemiology and psychiatric sciences}}
  \bibinfo{volume}{20}, \bibinfo{number}{3} (\bibinfo{year}{2011}),
  \bibinfo{pages}{231--233}.
\newblock


\bibitem[Craswell et~al\mbox{.}(2008)]%
        {craswell2008experimental}
\bibfield{author}{\bibinfo{person}{Nick Craswell}, \bibinfo{person}{Onno
  Zoeter}, \bibinfo{person}{Michael Taylor}, {and} \bibinfo{person}{Bill
  Ramsey}.} \bibinfo{year}{2008}\natexlab{}.
\newblock \showarticletitle{An experimental comparison of click position-bias
  models}. In \bibinfo{booktitle}{\emph{Proceedings of the 2008 international
  conference on web search and data mining}}. \bibinfo{pages}{87--94}.
\newblock


\bibitem[Danescu-Niculescu-Mizil et~al\mbox{.}(2010)]%
        {danescu2010competing}
\bibfield{author}{\bibinfo{person}{Cristian Danescu-Niculescu-Mizil},
  \bibinfo{person}{Andrei~Z Broder}, \bibinfo{person}{Evgeniy Gabrilovich},
  \bibinfo{person}{Vanja Josifovski}, {and} \bibinfo{person}{Bo Pang}.}
  \bibinfo{year}{2010}\natexlab{}.
\newblock \showarticletitle{Competing for users' attention: on the interplay
  between organic and sponsored search results}. In
  \bibinfo{booktitle}{\emph{Proceedings of the 19th international conference on
  World wide web}}. \bibinfo{pages}{291--300}.
\newblock


\bibitem[Draws et~al\mbox{.}(2021)]%
        {draws2021not}
\bibfield{author}{\bibinfo{person}{Tim Draws}, \bibinfo{person}{Nava Tintarev},
  \bibinfo{person}{Ujwal Gadiraju}, \bibinfo{person}{Alessandro Bozzon}, {and}
  \bibinfo{person}{Benjamin Timmermans}.} \bibinfo{year}{2021}\natexlab{}.
\newblock \showarticletitle{This Is Not What We Ordered: Exploring Why Biased
  Search Result Rankings Affect User Attitudes on Debated Topics}.
\newblock  (\bibinfo{year}{2021}), \bibinfo{pages}{295–305}.
\newblock
\showISBNx{9781450380379}
\urldef\tempurl%
\url{https://doi.org/10.1145/3404835.3462851}
\showURL{%
\tempurl}


\bibitem[Epstein and Robertson(2015)]%
        {epstein2015search}
\bibfield{author}{\bibinfo{person}{Robert Epstein} {and}
  \bibinfo{person}{Ronald~E Robertson}.} \bibinfo{year}{2015}\natexlab{}.
\newblock \showarticletitle{The search engine manipulation effect (SEME) and
  its possible impact on the outcomes of elections}.
\newblock \bibinfo{journal}{\emph{Proceedings of the National Academy of
  Sciences}} \bibinfo{volume}{112}, \bibinfo{number}{33}
  (\bibinfo{year}{2015}), \bibinfo{pages}{E4512--E4521}.
\newblock


\bibitem[Foulds et~al\mbox{.}(2021)]%
        {foulds2021investigating}
\bibfield{author}{\bibinfo{person}{Olivia Foulds}, \bibinfo{person}{Leif
  Azzopardi}, {and} \bibinfo{person}{Martin Halvey}.}
  \bibinfo{year}{2021}\natexlab{}.
\newblock \showarticletitle{Investigating the influence of ads on user search
  performance, behaviour, and experience during information seeking}. In
  \bibinfo{booktitle}{\emph{Proceedings of the 2021 Conference on Human
  Information Interaction and Retrieval}}. \bibinfo{pages}{107--117}.
\newblock


\bibitem[Gauzente(2010)]%
        {gauzente2010intention}
\bibfield{author}{\bibinfo{person}{Claire Gauzente}.}
  \bibinfo{year}{2010}\natexlab{}.
\newblock \showarticletitle{The intention to click on sponsored ads—A study
  of the role of prior knowledge and of consumer profile}.
\newblock \bibinfo{journal}{\emph{Journal of Retailing and Consumer Services}}
  \bibinfo{volume}{17}, \bibinfo{number}{6} (\bibinfo{year}{2010}),
  \bibinfo{pages}{457--463}.
\newblock


\bibitem[Gillies et~al\mbox{.}(2012)]%
        {gillies2012polyunsaturated}
\bibfield{author}{\bibinfo{person}{Donna Gillies}, \bibinfo{person}{John~KH
  Sinn}, \bibinfo{person}{Sagar~S Lad}, \bibinfo{person}{Matthew~J Leach},
  {and} \bibinfo{person}{Melissa~J Ross}.} \bibinfo{year}{2012}\natexlab{}.
\newblock \showarticletitle{Polyunsaturated fatty acids (PUFA) for attention
  deficit hyperactivity disorder (ADHD) in children and adolescents}.
\newblock \bibinfo{journal}{\emph{Cochrane Database of Systematic Reviews}}
  \bibinfo{number}{7} (\bibinfo{year}{2012}).
\newblock


\bibitem[Giraldo-Romero et~al\mbox{.}(2021)]%
        {giraldo2021influence}
\bibfield{author}{\bibinfo{person}{Yessica-Ileana Giraldo-Romero},
  \bibinfo{person}{Francisco Mu{\~n}oz-Leiva}, \bibinfo{person}{Elena
  Higueras-Castillo}, \bibinfo{person}{Francisco Li{\'e}bana-Cabanillas},
  {et~al\mbox{.}}} \bibinfo{year}{2021}\natexlab{}.
\newblock \showarticletitle{Influence of regulatory fit theory on persuasion
  from google ads: An eye tracking study}.
\newblock \bibinfo{journal}{\emph{Journal of Theoretical and Applied Electronic
  Commerce Research}} \bibinfo{volume}{16}, \bibinfo{number}{5}
  (\bibinfo{year}{2021}), \bibinfo{pages}{1165--1185}.
\newblock


\bibitem[Harpin et~al\mbox{.}(2016)]%
        {harpin2016long}
\bibfield{author}{\bibinfo{person}{V Harpin}, \bibinfo{person}{L Mazzone},
  \bibinfo{person}{JP Raynaud}, \bibinfo{person}{J Kahle}, {and}
  \bibinfo{person}{P Hodgkins}.} \bibinfo{year}{2016}\natexlab{}.
\newblock \showarticletitle{Long-term outcomes of ADHD: a systematic review of
  self-esteem and social function}.
\newblock \bibinfo{journal}{\emph{Journal of attention disorders}}
  \bibinfo{volume}{20}, \bibinfo{number}{4} (\bibinfo{year}{2016}),
  \bibinfo{pages}{295--305}.
\newblock


\bibitem[Hashavit et~al\mbox{.}(2021)]%
        {hashavit2021understanding}
\bibfield{author}{\bibinfo{person}{Anat Hashavit}, \bibinfo{person}{Hongning
  Wang}, \bibinfo{person}{Raz Lin}, \bibinfo{person}{Tamar Stern}, {and}
  \bibinfo{person}{Sarit Kraus}.} \bibinfo{year}{2021}\natexlab{}.
\newblock \showarticletitle{Understanding and Mitigating Bias in Online Health
  Search}.
\newblock  (\bibinfo{year}{2021}), \bibinfo{pages}{265–274}.
\newblock
\showISBNx{9781450380379}
\urldef\tempurl%
\url{https://doi.org/10.1145/3404835.3462930}
\showURL{%
\tempurl}


\bibitem[Heirs and Dean(2007)]%
        {heirs2007homeopathy}
\bibfield{author}{\bibinfo{person}{Morag Heirs} {and}
  \bibinfo{person}{Mike~Emmans Dean}.} \bibinfo{year}{2007}\natexlab{}.
\newblock \showarticletitle{Homeopathy for attention deficit/hyperactivity
  disorder or hyperkinetic disorder}.
\newblock \bibinfo{journal}{\emph{Cochrane database of systematic reviews}}
  \bibinfo{number}{4} (\bibinfo{year}{2007}).
\newblock


\bibitem[Herxheimer and Petrie(2002)]%
        {herxheimer2002melatonin}
\bibfield{author}{\bibinfo{person}{Andrew Herxheimer} {and}
  \bibinfo{person}{Keith~J Petrie}.} \bibinfo{year}{2002}\natexlab{}.
\newblock \showarticletitle{Melatonin for the prevention and treatment of jet
  lag}.
\newblock \bibinfo{journal}{\emph{Cochrane Database of Systematic Reviews}}
  \bibinfo{number}{2} (\bibinfo{year}{2002}).
\newblock


\bibitem[Hilton et~al\mbox{.}(2013)]%
        {hilton2013ginkgo}
\bibfield{author}{\bibinfo{person}{Malcolm~P Hilton},
  \bibinfo{person}{Eleanor~F Zimmermann}, {and} \bibinfo{person}{William~T
  Hunt}.} \bibinfo{year}{2013}\natexlab{}.
\newblock \showarticletitle{Ginkgo biloba for tinnitus}.
\newblock \bibinfo{journal}{\emph{Cochrane Database of Systematic Reviews}}
  \bibinfo{number}{3} (\bibinfo{year}{2013}).
\newblock


\bibitem[Jansen and Resnick(2006)]%
        {jansen2006examination}
\bibfield{author}{\bibinfo{person}{Bernard~J Jansen} {and}
  \bibinfo{person}{Marc Resnick}.} \bibinfo{year}{2006}\natexlab{}.
\newblock \showarticletitle{An examination of searcher's perceptions of
  nonsponsored and sponsored links during ecommerce Web searching}.
\newblock \bibinfo{journal}{\emph{Journal of the American Society for
  information Science and Technology}} \bibinfo{volume}{57},
  \bibinfo{number}{14} (\bibinfo{year}{2006}), \bibinfo{pages}{1949--1961}.
\newblock


\bibitem[Joachims et~al\mbox{.}(2017)]%
        {joachims2017accurately}
\bibfield{author}{\bibinfo{person}{Thorsten Joachims}, \bibinfo{person}{Laura
  Granka}, \bibinfo{person}{Bing Pan}, \bibinfo{person}{Helene Hembrooke},
  {and} \bibinfo{person}{Geri Gay}.} \bibinfo{year}{2017}\natexlab{}.
\newblock \showarticletitle{Accurately interpreting clickthrough data as
  implicit feedback}. In \bibinfo{booktitle}{\emph{ACM SIGIR Forum}},
  Vol.~\bibinfo{volume}{51}. Acm New York, NY, USA, \bibinfo{pages}{4--11}.
\newblock


\bibitem[Keane et~al\mbox{.}(2008)]%
        {keane2008people}
\bibfield{author}{\bibinfo{person}{Mark~T Keane}, \bibinfo{person}{Maeve
  O'Brien}, {and} \bibinfo{person}{Barry Smyth}.}
  \bibinfo{year}{2008}\natexlab{}.
\newblock \showarticletitle{Are people biased in their use of search engines?}
\newblock \bibinfo{journal}{\emph{Commun. ACM}} \bibinfo{volume}{51},
  \bibinfo{number}{2} (\bibinfo{year}{2008}), \bibinfo{pages}{49--52}.
\newblock


\bibitem[Lagun et~al\mbox{.}(2016)]%
        {lagun2016understanding}
\bibfield{author}{\bibinfo{person}{Dmitry Lagun}, \bibinfo{person}{Donal
  McMahon}, {and} \bibinfo{person}{Vidhya Navalpakkam}.}
  \bibinfo{year}{2016}\natexlab{}.
\newblock \showarticletitle{Understanding mobile searcher attention with rich
  ad formats}. In \bibinfo{booktitle}{\emph{Proceedings of the 25th ACM
  International on Conference on Information and Knowledge Management}}.
  \bibinfo{pages}{599--608}.
\newblock


\bibitem[Lewandowski(2017)]%
        {lewandowski2017users}
\bibfield{author}{\bibinfo{person}{Dirk Lewandowski}.}
  \bibinfo{year}{2017}\natexlab{}.
\newblock \showarticletitle{Users' understanding of search engine
  advertisements}.
\newblock \bibinfo{journal}{\emph{Journal of Information Science Theory and
  Practice}} \bibinfo{volume}{5}, \bibinfo{number}{4} (\bibinfo{year}{2017}),
  \bibinfo{pages}{6--25}.
\newblock


\bibitem[Lewandowski et~al\mbox{.}(2018)]%
        {lewandowski2018empirical}
\bibfield{author}{\bibinfo{person}{Dirk Lewandowski},
  \bibinfo{person}{Friederike Kerkmann}, \bibinfo{person}{Sandra R{\"u}mmele},
  {and} \bibinfo{person}{Sebastian S{\"u}nkler}.}
  \bibinfo{year}{2018}\natexlab{}.
\newblock \showarticletitle{An empirical investigation on search engine ad
  disclosure}.
\newblock \bibinfo{journal}{\emph{Journal of the Association for Information
  Science and Technology}} \bibinfo{volume}{69}, \bibinfo{number}{3}
  (\bibinfo{year}{2018}), \bibinfo{pages}{420--437}.
\newblock


\bibitem[Li(2019)]%
        {li2019user}
\bibfield{author}{\bibinfo{person}{Yujie Li}.} \bibinfo{year}{2019}\natexlab{}.
\newblock \showarticletitle{User perception affects search engine advertising
  avoidance: Moderating role of user characteristics}.
\newblock \bibinfo{journal}{\emph{Social Behavior and Personality: an
  international journal}} \bibinfo{volume}{47}, \bibinfo{number}{4}
  (\bibinfo{year}{2019}), \bibinfo{pages}{1--12}.
\newblock


\bibitem[Lu et~al\mbox{.}(2017)]%
        {lu2017sponsored}
\bibfield{author}{\bibinfo{person}{Yan Lu}, \bibinfo{person}{Michael Chau},
  {and} \bibinfo{person}{Patrick~YK Chau}.} \bibinfo{year}{2017}\natexlab{}.
\newblock \showarticletitle{Are sponsored links effective? Investigating the
  impact of trust in search engine advertising}.
\newblock \bibinfo{journal}{\emph{ACM Transactions on Management Information
  Systems (TMIS)}} \bibinfo{volume}{7}, \bibinfo{number}{4}
  (\bibinfo{year}{2017}), \bibinfo{pages}{1--33}.
\newblock


\bibitem[O’Brien and Keane(2006)]%
        {o2006modeling}
\bibfield{author}{\bibinfo{person}{Maeve O’Brien} {and}
  \bibinfo{person}{Mark~T Keane}.} \bibinfo{year}{2006}\natexlab{}.
\newblock \showarticletitle{Modeling result-list searching in the World Wide
  Web: The role of relevance topologies and trust bias}. In
  \bibinfo{booktitle}{\emph{Proceedings of the 28th annual conference of the
  cognitive science society}}, Vol.~\bibinfo{volume}{28}. Citeseer,
  \bibinfo{pages}{1881--1886}.
\newblock


\bibitem[Pan et~al\mbox{.}(2007)]%
        {pan2007google}
\bibfield{author}{\bibinfo{person}{Bing Pan}, \bibinfo{person}{Helene
  Hembrooke}, \bibinfo{person}{Thorsten Joachims}, \bibinfo{person}{Lori
  Lorigo}, \bibinfo{person}{Geri Gay}, {and} \bibinfo{person}{Laura Granka}.}
  \bibinfo{year}{2007}\natexlab{}.
\newblock \showarticletitle{In Google we trust: Users’ decisions on rank,
  position, and relevance}.
\newblock \bibinfo{journal}{\emph{Journal of computer-mediated communication}}
  \bibinfo{volume}{12}, \bibinfo{number}{3} (\bibinfo{year}{2007}),
  \bibinfo{pages}{801--823}.
\newblock


\bibitem[Phillips et~al\mbox{.}(2013)]%
        {phillips2013ads}
\bibfield{author}{\bibinfo{person}{Adrienne~Hall Phillips},
  \bibinfo{person}{Ruijiao Yang}, {and} \bibinfo{person}{Soussan Djamasbi}.}
  \bibinfo{year}{2013}\natexlab{}.
\newblock \showarticletitle{Do ads matter? An exploration of web search
  behavior, visual hierarchy, and search engine results pages}. In
  \bibinfo{booktitle}{\emph{2013 46th Hawaii International Conference on System
  Sciences}}. IEEE, \bibinfo{pages}{1563--1568}.
\newblock


\bibitem[Pogacar et~al\mbox{.}(2017)]%
        {pogacar2017positive}
\bibfield{author}{\bibinfo{person}{Frances~A Pogacar}, \bibinfo{person}{Amira
  Ghenai}, \bibinfo{person}{Mark~D Smucker}, {and} \bibinfo{person}{Charles~LA
  Clarke}.} \bibinfo{year}{2017}\natexlab{}.
\newblock \showarticletitle{The positive and negative influence of search
  results on people's decisions about the efficacy of medical treatments}. In
  \bibinfo{booktitle}{\emph{Proceedings of the ACM SIGIR International
  Conference on Theory of Information Retrieval}}. \bibinfo{pages}{209--216}.
\newblock


\bibitem[Schulthei{\ss} and Lewandowski(2021)]%
        {schultheiss2021users}
\bibfield{author}{\bibinfo{person}{Sebastian Schulthei{\ss}} {and}
  \bibinfo{person}{Dirk Lewandowski}.} \bibinfo{year}{2021}\natexlab{}.
\newblock \showarticletitle{How users' knowledge of advertisements influences
  their viewing and selection behavior in search engines}.
\newblock \bibinfo{journal}{\emph{Journal of the Association for Information
  Science and Technology}} \bibinfo{volume}{72}, \bibinfo{number}{3}
  (\bibinfo{year}{2021}), \bibinfo{pages}{285--301}.
\newblock


\bibitem[Varian et~al\mbox{.}(2006)]%
        {varian2006economics}
\bibfield{author}{\bibinfo{person}{Hal~R Varian} {et~al\mbox{.}}}
  \bibinfo{year}{2006}\natexlab{}.
\newblock \showarticletitle{The economics of internet search}.
\newblock \bibinfo{journal}{\emph{Rivista di politica economica}}
  \bibinfo{volume}{96}, \bibinfo{number}{11/12} (\bibinfo{year}{2006}),
  \bibinfo{pages}{8}.
\newblock


\bibitem[White(2013)]%
        {white2013beliefs}
\bibfield{author}{\bibinfo{person}{Ryen White}.}
  \bibinfo{year}{2013}\natexlab{}.
\newblock \showarticletitle{Beliefs and biases in web search}. In
  \bibinfo{booktitle}{\emph{Proceedings of the 36th international ACM SIGIR
  conference on Research and development in information retrieval}}.
  \bibinfo{pages}{3--12}.
\newblock


\bibitem[White(2014)]%
        {white2014belief}
\bibfield{author}{\bibinfo{person}{Ryen~W White}.}
  \bibinfo{year}{2014}\natexlab{}.
\newblock \showarticletitle{Belief dynamics in Web search}.
\newblock \bibinfo{journal}{\emph{Journal of the Association for Information
  Science and Technology}} \bibinfo{volume}{65}, \bibinfo{number}{11}
  (\bibinfo{year}{2014}), \bibinfo{pages}{2165--2178}.
\newblock


\bibitem[White and Hassan(2014)]%
        {white2014content}
\bibfield{author}{\bibinfo{person}{Ryen~W White} {and} \bibinfo{person}{Ahmed
  Hassan}.} \bibinfo{year}{2014}\natexlab{}.
\newblock \showarticletitle{Content bias in online health search}.
\newblock \bibinfo{journal}{\emph{ACM Transactions on the Web (TWEB)}}
  \bibinfo{volume}{8}, \bibinfo{number}{4} (\bibinfo{year}{2014}),
  \bibinfo{pages}{1--33}.
\newblock


\bibitem[Yue et~al\mbox{.}(2010)]%
        {yue2010beyond}
\bibfield{author}{\bibinfo{person}{Yisong Yue}, \bibinfo{person}{Rajan Patel},
  {and} \bibinfo{person}{Hein Roehrig}.} \bibinfo{year}{2010}\natexlab{}.
\newblock \showarticletitle{Beyond position bias: Examining result
  attractiveness as a source of presentation bias in clickthrough data}. In
  \bibinfo{booktitle}{\emph{Proceedings of the 19th international conference on
  World wide web}}. \bibinfo{pages}{1011--1018}.
\newblock


\bibitem[Zenetti et~al\mbox{.}(2014)]%
        {searchAdsEffect}
\bibfield{author}{\bibinfo{person}{German Zenetti}, \bibinfo{person}{Tammo
  H.~A. Bijmolt}, \bibinfo{person}{Peter S.~H. Leeflang}, {and}
  \bibinfo{person}{Daniel Klapper}.} \bibinfo{year}{2014}\natexlab{}.
\newblock \showarticletitle{Search Engine Advertising Effectiveness in a
  Multimedia Campaign}.
\newblock \bibinfo{journal}{\emph{International Journal of Electronic
  Commerce}} \bibinfo{volume}{18}, \bibinfo{number}{3} (\bibinfo{year}{2014}),
  \bibinfo{pages}{7--38}.
\newblock


\end{thebibliography}

\end{document}